**Open Access effect on uncitedness: A large-scale study controlling by discipline, source type and visibility**

Pablo Dorta-González *

Universidad de Las Palmas de Gran Canaria, TiDES Research Institute, Campus de Tafira, 35017 Las Palmas de Gran Canaria, Spain. *E-mail*: pablo.dorta@ulpgc.es

* Corresponding author

Rafael Suárez-Vega

Universidad de Las Palmas de Gran Canaria, TiDES Research Institute, Campus de Tafira, 35017 Las Palmas de Gran Canaria, Spain. *E-mail*: rafael.suarez@ulpgc.es

María Isabel Dorta-González

Universidad de La Laguna, Departamento de Ingeniería Informática y de Sistemas, Avenida Astrofísico Francisco Sánchez s/n, 38271 La Laguna, Spain. *E-mail*: isadorta@ull.es

**Abstract**

There are many factors that affect the probability of being uncited during the first years after publication. In this study, we analyze three of these factors for journals, conference proceedings and book series: the field (in 316 subject categories of the Scopus database), the access modality (open access vs. paywalled), and the visibility of the source (through the percentile of the average impact in the subject category). We quantify the effect of these factors on the probability of being uncited. This probability is measured through the percentage of uncited documents in the serial sources of the Scopus database at about two years after publication. As a main result, we do not find any strong correlation between open access and uncitedness. Within the group of most cited journals (Q1 and top 10%), open access journals generally have somewhat lower uncited rates. However, in the intermediate quartiles (Q2 and Q3) almost no differences are observed, while for Q4 the uncited rate is again somewhat lower in the case of the OA group. This is important because it provides new evidence in the debate about open access citation advantage.

**1. Introduction**

Uncited research has received wide attention in the literature because of the relevance of the phenomenon of uncitedness to research policy. Some authors claim that uncited publications have no influence on future research and might be a waste of resources (Van Noorden, 2017). However, the phenomenon of uncitedness needs to be better understood and characterized before it can be used in an assessment of productivity.

Authors have focused on estimations of the frequency and characterization of uncited papers. Some authors have even used a control group of highly cited papers (Yamashita and Yoshinaga, 2014; Kamat, 2018). The measure of uncitedness has been recently reviewed by Nicolaisen and Frandsen (2019). It has been observed that the uncitedness ratio (the fraction of uncited papers in a collection) strongly depends on the observation time window, the discipline, and the document type (Van Leeuwen and Moed, 2005; Wallace, Larivière and Gingras, 2009; Thelwall, 2016).

Although citedness is typically the focus of bibliometric analyses, empirical studies are affected by uncited publications. Thelwall (2016) showed there was a correlation between the uncitedness ratio and the shape of the citation distribution. Some authors have provided uncited rates for journals with their impact factor, which is the mean number of citations per paper in the first two years after publication (Van Leeuwen and Moed, 2005; Hsu and Huang, 2012; Burrell, 2013; Egghe, 2013).

Uncitedness refers to the status of academic publications, authors or fields that do not receive any citation within a time window (Liang, Zhong and Rousseau, 2015). Consequently, the concept of uncitedness depends to a large extent on the length of the time window (Hu and Wu, 2014). However, the fact that a publication is currently uncited does not mean that it will never be cited (Van Raan, 2015; Ho and Hartley, 2017).

Uncitedness can be analyzed at author level. Using Nobel laureates and Fields medalists, Egghe, Guns and Rousseau (2011) found uncitedness rates of over 10%. Heneberg (2013), however, argued that the high uncited ratios were motivated by the inclusion of uncitable document types in the analyses and, focusing only on papers and reviews, reported uncited rates below 1%.

Uncitedness can also be analyzed at field level (Mavrogenis et al., 2018; Rosenkrantz, Chung and Duszak, 2018). Liang, Zhong and Rousseau (2015) found that low numbers of pages, references, and authors per paper were associated with uncitedness in library and information science. Lou and He (2015) found a weak negative correlation between affiliation reputation and uncitedness in six subject areas.

Previous studies have shown the need to analyze uncitedness with the inclusion of two explanatory factors: discipline and document typology. However, no studies have been published that analyze the possible influence on the uncited rate of access modality (open access (OA) and paywalled). Open access citedness has recently been studied in the case of journals (Dorta-González, González-Betancor and Dorta-González, 2017; Dorta-González and Santana-Jiménez, 2018; González-Betancor and Dorta-González, 2019). However, as previously stated, uncitedness is a different phenomenon that needs to be better understood and characterized before it can be used in an assessment of productivity.

With the above in mind, in this paper we conduct a large-scale analysis of the influence on the uncited rate of field, document type, access modality, and journal visibility. We use the percentage of uncited documents in a serial collection (journals, conference proceedings, and book series) as a better measure of the frequency of low-impact documents within that collection than the CiteScore ranking which measures the average impact of all published documents in the same collection.

## 2. Methodology

The following information for each serial title (excluding trade journals as our interest is only the analysis of original research) was downloaded from the Scopus database (February 2020) to an Excel file: the number of documents published in the years 2015 to 2017, and the citations received by the same documents in the year 2018. The following variables were also obtained from this database:

-Source typology: Qualitative (journal, conference proceeding, book series)

-Access modality: Qualitative (OA, paywalled)

-Subject category: Qualitative (316 disciplines with 5 or more journals). Note that each serial title is in one or more discipline. The list of disciplines can be consulted in Appendix A.

-Subject area: Qualitative (27 in total). Note that each serial title is in one or more subject areas. The list of subject areas and their categories can be consulted in Appendix A.

-Branch of knowledge: Qualitative (Health Sciences; Life Sciences; Physical Sciences; Social Sciences & Humanities).

-CiteScore (citations per document): Continuous (range 0-160.19). CiteScore calculates the average number of citations received in a calendar year by all items published in a serial title in the preceding three years. In this study, CiteScore calculates the average number of citations received in year 2018 by all items published in that serial title in the years 2015, 2016, and 2017.

-Percentile in the subject category: Integer (range 0-99). The CiteScore percentile indicates the relative impact of a serial title in its subject category. Each subject category is divided into 100 equal-sized percentiles based on the number of serial titles, and a serial title is assigned to a percentile based on its CiteScore. For example, a serial title that has a CiteScore percentile of 90% is ranked according to CiteScore as higher than 90% of the serial titles within that category (i.e. in the top 10% of the subject category). Note that a percentile of 75% or above is considered as being in the first quartile (Q1), between 50% and 75% in the second quartile (Q2), between 25% and 50% in the third quartile (Q3), and below 25% in the fourth quartile (Q4).

-Uncited rate: Integer (range 0-100). This indicates the consistency with which the documents in a serial title are uncited. In the context of CiteScore, the uncited rate is the proportion of documents considered in the denominator of the CiteScore calculation that have not received any citations in the CiteScore numerator. Note that in the case of some documents three full years will have elapsed since publication whereas for others only one full year will have passed. Therefore, the uncited rate is a measure that corresponds on average to a date two years after publication.

## 3. Results

*3.1 Field and access modality effects for journals*

The mean of the uncited rate for the 316 disciplines (Scopus subject categories) with five or more journals can be seen in Appendix A. This information is disaggregated by access modality (OA vs. paywalled). Note that there are no OA journals in 9 of the disciplines. In half of the disciplines, the average uncited rate for journals is higher than 47% at two years after publication (median = 47). The uncited rate ranges from a minimum value of 23% to a maximum of 89%. Although there is a high degree of variability in uncitedness (standard deviation = 11.87), the average uncited rate for journals is 48.96% at two years after publication (see Table 1). This same information but aggregated by subject area and access modality is shown in Figure 1.

[Figure 1 about here]

Journals with uncited rates below 25% are found in only two disciplines: Cellular and Molecular Neuroscience (23%) in the area of Neuroscience, and Catalysis (24%) in the area of Chemical Engineering. Most disciplines with lower uncitedness correspond to the branches of Life Sciences (especially in Chemistry and Materials) and Health Sciences. Of the 37 disciplines with ratios below 35%, 12 are in in Biochemistry, Genetics and Molecular Biology; 9 in Neuroscience; 4 in Chemistry; 4 in Chemical Engineering; 3 in Immunology and Microbiology; 2 in Environmental Science; 1 in Materials Science; 1 in Medicine; and 1 in Pharmacology, Toxicology and Pharmaceutics.

Conversely, uncited rates above 85% are obtained in four disciplines in Arts and Humanities: Literature and Literary Theory (89%), Classics (87%), Visual Arts and Performing Arts (87%), and Religious Studies (85%). A total of 12 out of 14 disciplines in Arts and Humanities have uncitedness rates above 73%. The other two are History and Philosophy of Science (68%), and Arts and Humanities -miscellaneous- (52%). Above 70% there are only two disciplines from other areas: Pharmacology (81%) in Nursing, and Cultural Studies (78%) in Social Sciences.

The central tendency and variability measures for the average uncited rate, disaggregated by access modality, are shown in Table 1. In half of the 316 disciplines the average uncited rate for OA journals is higher than 51% at two years after publication (median = 51). The uncited rate ranges from a minimum value of 18% to a maximum of 93%. Although there is a high degree of variability in uncitedness (standard deviation = 13.69), the average uncited rate for journals is 51.32% at two years after publication. In the case of paywalled journals, in half of the disciplines the average uncited rate for this type of access is higher than 46.5% at two years after publication. This is 4.5 percentage points less than the median in the case of OA journals. The uncited rate ranges from a minimum value of 23% to a maximum of 89%. This is 9 percentage points less in the variation range in comparison with the OA group. Although there is also a high degree of variability in uncitedness (standard deviation = 12.22), the average uncited rate for the paywalled journals is 48.57% at two years after publication. This is three percentage points below uncitedness in the OA group.

[Table 1 about here]

The distribution of the mean uncited rate in the disciplines is shown in Figure 2. The information is disaggregated by access modality. The box diagram shows for each access modality the atypical values, the maximum and minimum values, as well as the median and quartiles. The mean is also represented by a cross.

[Figure 2 about here]

As can be seen in Figure 2, the uncited rates in the group of OA journals are generally higher than those in the group of paywalled journals. This is because the boxes in the data distribution are slightly displaced upwards in the diagram. Both the median and the cross that represents the mean are also clearly higher in the case of OA journals. However, there are two outliers that correspond to disciplines in which the uncited rate is significantly lower in the case of OA journals compared to paywalled ones. These are the points represented at an uncited rate of around 20%: Cellular and Molecular Neuroscience in the area of Neuroscience, and Immunology and Microbiology -miscellaneous- in the area of Immunology and Microbiology. However, in the latter discipline there are only 5 journals, two of which are OA with an uncited rate of 18% in comparison with 60% for the three paywalled ones.

The differences between access modality by discipline can be seen much better in Figure 3. This representation shows the difference in the mean uncited rate (OA minus paywalled) by discipline. Note that Appendix A also shows this difference with colors that represent the magnitude of the subtraction.

[Figure 3 about here]

In Figure 3, it can be observed that in a majority of the disciplines the uncited rate is higher in the group of OA journals (and therefore the subtraction is positive). Specifically, in 206 out of 316 disciplines, the difference is positive (65% of cases), which represents a greater proportion of cases in which the uncited rate is higher within OA journals. In another 93 of the 316 disciplines, the difference is negative (29% of cases), corresponding to those disciplines in which the uncited rate is higher in paywalled journals. In the other 17 disciplines, the uncited rate either coincides for the two access modalities or there is no OA journal with which to compare.

The maximum values for the difference in the positive sign group are around 25 percentage points, but this difference exceeds 20 percentage points in only 11 cases. These disciplines correspond to Arts and Humanities -miscellaneous- (27 points of difference for 20 OA journals), Periodontics (27 points for 5 OA journals), Logic (25 points and 3 OA journals), Tourism, Leisure and Hospitality Management (24 points for 14 OA journals), Environmental Chemistry (24 points for 8 OA journals), Life-span and Life-course Studies (24 points and 5 OA journals), Oral Surgery (23 points for 12 OA journals), Nursing -miscellaneous- (21 points but only 1 OA journal), Social Sciences -miscellaneous- (21 points for 35 OA journals), Numerical Analysis (20 points for 7 OA journals), and Developmental and Educational Psychology (20 points for 20 OA journals).

Although there is a smaller proportion of disciplines for which the uncited rate is higher within the paywalled group, and therefore the subtraction is negative, the magnitude of these differences is larger in some cases. In 11 disciplines the difference exceeds 20 percentage points, but note that in 5 of these cases the difference exceeds 30 percentage points, and even 40 points in 2 cases. However, the cases where this difference is greater corresponds to disciplines with few OA journals with which to compare (two or even only one). These disciplines are Immunology and Microbiology -miscellaneous- (42 points of difference for 2 OA journals), Podiatry (40 points but only 1 OA journal), Fundamentals and skills (32 points for only 1 OA journal), Maternity and Midwifery (31 points for 2 OA journals), Pharmacology, Toxicology and Pharmaceutics -miscellaneous- (31 points for 4 OA journals), Biochemistry, Genetics and Molecular Biology -miscellaneous- (22 points for 10 OA journals), Complementary and Alternative Medicine (21 points for 19 OA journals), Family Practice (21 points for 19 OA journals), Issues, Ethics and Legal Aspects (21 points for 4 OA journals), Statistical and Nonlinear Physics (21 points for 3 OA journals), and Small Animals (21 points of difference but for only one journal).

The above can be seen graphically in Figure 4. This box diagram shows the distribution for the difference of uncited rates for journals in the disciplines of Figure 3. It can be seen that the boxes are slightly displaced towards the positive part of the axis, with both the median and the mean clearly located above zero. Outliers correspond to the disciplines mentioned in the previous paragraph.

[Figure 4 about here]

*3.2 Visibility and access effects for journals, conference proceedings and book series*

A box diagram for the distribution of uncited rate for journals by access modality and average impact ranking (CiteScore percentile) is shown in Figure 5. Although there are outliers in all quartiles, even within the group of journals located in the top 10% of the most cited, the following results and trends can be clearly observed. In general, the uncited rate decreases as journal average impact increases. As for access modality, when comparing journals of similar impact, the differences that were observed previously for aggregated data of visibility, are reduced. It even seems that the amplitude of the variation range excluding the outliers is somewhat less in the case of OA journals. Within the group of the most cited journals (Q1 and top 10%), OA journals generally have somewhat lower uncited rates. However, in the intermediate quartiles (Q2 and Q3) almost no differences are observed, while for Q4 the uncited rate is again somewhat lower in the case of the OA group.

[Figure 5 about here]

It is also surprising that within the group of top 10% of the most cited journals, some have uncited rates greater than 50% at two years after publication, and even 80% for some paywalled journals. These journals correspond to the field of Humanities as previously indicated. In contrast, within the group of journals with the lowest average impact (Q4) some have uncited rates of less than 40%.

The box diagram for the distribution of uncited rates in the case of conference proceedings and book series is shown in Figure 6. Note that there are no OA conference proceedings. There is greater variability for the paywalled conference proceedings than in the case of journals. The uncited rate ranges from 5% to 100%, much wider than that observed in the case of journals (see Figure 2). In half of the proceedings, the uncited rate is greater than 60%. This is significantly higher than that obtained for journals, 45% and 50% for OA and paywalled, respectively (see Figure 2).

[Figure 6 about here]

In the case of book series, the first thing that stands out is a lower tail for the distribution which is much longer in the case of paywalled book series, indicating that, surprisingly, OA does not guarantee better visibility and impact. In general, OA books have higher uncited rates than paywalled ones.

Note that uncited rates for book series are much higher than for journals and conference proceedings. However, in the areas in which the use of books is most widespread as a channel of communication for the results of the research (Humanities), the maturation time of citations is much longer, and therefore a citation window of two years after publication may be too short to measure the real impact in the medium and long term. This is quite different to the case of journals where, in the vast majority of cases, the maximum citation distribution is attained between two and three years after publication.

A box diagram for the distribution of CiteScore for journals (by access modality and visibility) is shown in Figure 7. In this case, the outliers are not shown to allow better visualization of the vast majority of cases. It can be observed that in the group of journals with the greatest impact (Q1 and Top 10%) both the mean and the median are close between access modalities, although it is true that the upper tail is somewhat longer in the case of the paywalled journals. In the intermediate quartiles (Q2 and Q3) there are practically no differences in the measures of central tendency and variability, with no significant differences being observed between the two journal types. In contrast, within the group of journals with the lowest impact (Q4) the measures of central tendency are superior in the case of OA. This indicates that OA facilitates citations for journals with worse visibility, which usually corresponds to those journals for which many institutions do not have a subscription.

[Figure 7 about here]

*3.3 Correlation between uncited rate and average impact (CiteScore and Percentile) according to field and access modality*

A scatter plot between the uncited rate and the percentile is shown in Figure 8. Six disciplines (subject categories) with different document types are used as case studies. In the case of the journals, the four disciplines considered as case studies to analyze the possible correlation between uncited rate and percentile correspond to subject categories of very different sizes in terms of the number of journals, and from the four different branches of knowledge. These disciplines are: Medicine (A), with 202 journals from the Health Sciences; Plant Science (B), with 398 journals from the Life Sciences; Physics and Astronomy (C), with 41 journals from the Physical Sciences; and Economics and Econometrics (D), with 579 journals from the Social Sciences & Humanities.

[Figure 8 about here]

To analyze the correlation in the case of conference proceedings, the subject category Electrical and Electronic Engineering (E) was selected. This is an example of a discipline in which this means of scientific communication is widespread (39 proceedings). Finally, the discipline History (F) was selected to analyse the correlation in the book series typology. This is also an example of a discipline where this means of scientific communication is widespread (103 book series).

In the six case studies, a strong negative correlation was observed between the uncited rate and the percentile. The higher the percentile of the serial title, the lower the uncited rate. The correlation coefficient is close to 0.95 in the case of journals, 0.97 in the case of conference proceedings, and falls to 0.90 in the case of book series. In general, a greater dispersion in the points cloud (worse fit) is observed among the serial titles in the highest percentile (Q1).

Diagrams A, B and D show points of journals located in the first positions of the ranking by average impact (around 99th percentile) with uncited rates above 30%. This is due to one or more punctual successes (papers abnormally highly cited) that considerably increase the average impact of that journal. Conversely, there are some journals in these disciplines in which most papers receive no citation (uncited rate of 100%) two years after publication. These cases correspond to the journals located in the first percentiles. Note that the correlation is practically linear, although the fit is somewhat improved with second degree polynomials, and in some cases with a third degree polynomial (see diagram D).

In the case of book series (Diagram F), the slope of the curve is less pronounced than in the case of journals and conference proceedings. The points cloud moves to the top of the quadrant. This means that in all cases except one, the book series in History show uncited rates above 50% two years after publication. However, as previously stated, in the area in which the use of books is most widespread (Humanities), the maturation time of citations is much longer and two years may be too short a time to measure the real impact in the medium and long term.

The scatter plot between uncited rate and percentile in the case of OA journals is shown in Appendix B (see Figure B.1). Note that there are no OA conference proceedings in Electrical and Electronic Engineering, and only two OA book series in History. For the journals, in these case studies no significant differences are observed in relation to the total group of journals. The only appreciable difference is that it improves the fit of the regression curve within the group of journals with the greatest impact (Q1).

A bubble diagram for the source size (total documents) in the same six disciplines used as case studies is shown in Figure 9. The size of the bubble is proportional to the number of published documents in the serial title, and the coordinates are the uncited rate and percentile. The size of the serial title does not appear to be influencing either the uncited rate or the percentile. Journals of similar size are distributed uniformly across the entire points cloud. There are specific cases of serial titles that are much larger than the average, but these cases appear in some disciplines at the top of the curve and in others at the bottom, and no trend can be concluded from the data in this regard.

[Figure 9 about here]

A scatter plot between the uncited rate and the CiteScore is shown in Figure 10. The same six disciplines are used as case studies. A strong non-linear relationship is observed. The higher the average impact of the serial title, the lower the uncited rate. This has already been observed previously (see Figures 8 and 9). However, it can now be seen that this relationship is convex. This means that at the beginning of the curve (small CiteScore values), an increase in the number of citations (and therefore in the average impact) causes greater reductions in the uncited rate compared to the end of the curve (high CiteScore values). Furthermore, it is possible to observe the important differences that exist in the range of variation of the average impact in the serial titles (CiteScore). Most of the points present an average impact of less than 5 points in the case of journals, less than 2 points in the case of conference proceedings, and less than 0.5 points in the case of book series.

[Figure 10 about here]

Finally, the same scatter plot between uncited rate and CiteScore for the OA journals in the four disciplines used as case studies for journals is shown in Appendix B (see Figure B.2). Remember that there are no OA conference proceedings in Electrical and Electronic Engineering, and only two OA book series in History. However, no significant differences are observed (in relation to the shape of the regression curve) in these case studies with respect to the total group. The only appreciable difference is that it improves the fit of the curve within the group of journals with the greatest impact (at least in the case of Medicine and Plant Science).

**Discussion and Conclusions**

The phenomenon of uncitedness is relevant in research policy. In the past, authors have focused on estimation of the frequency and the characterization of uncited documents. However, uncitedness needs to be better understood and characterized before it can be used in an assessment of productivity. With this in mind, we conducted a large-scale analysis of uncitedness focusing on the influence of field, document typology, access modality, and source visibility.

There is a high degree of variability in uncitedness for the case of journals, but in half of the 316 disciplines considered in the present study the average uncited rate is higher than 47% at two years after publication. The disciplines with the highest uncited rates correspond to Humanities, while the lowest uncitedness is observed in Life Sciences (especially in Chemistry and Materials) and Health Sciences. As for access modality, the uncited rates in the group of OA journals are generally higher than those in the group of paywalled ones, at least when it is not distinguished by journal visibility. Specifically, in 65% of disciplines the uncited rate is higher within the OA journals, while in 29% of cases uncitedness is higher in the paywalled ones.

However, if disaggregated by journal visibility, the results are somewhat different. In general, the uncited rate decreases as journal average impact increases. Moreover, when comparing journals of similar impact, the differences that were observed previously for the aggregated data are reduced.

With respect to the correlation between uncitedness and the so-called "OA citation advantage", some considerations can be made. We do not find any strong correlation between OA and uncitedness. This is important because it provides new evidence in the debate about OA citation advantage. Within the group of most cited journals (Q1 and top 10%), OA journals generally have somewhat lower uncited rates. This could be due to the OA citation advantage effect. Papers published in the most widely distributed journals receive more citations when they are published openly. This OA citation advantage effect would also reduce the uncited rate in this group of top journals. This is because a part of the papers that would not receive any citation within a paywalled journal could now receive some citations when they are distributed in OA.

Something similar is observed within the journals with lowest visibility. In the last quartile (Q4), uncitedness is again somewhat lower in the case of the OA journals. This group corresponds to those journals for which many institutions do not have a subscription, and therefore OA facilitates visibility and impact. Unfortunately, institutions that cannot maintain full subscriptions to publishers first drop the subscription to this group of least read and cited journals.

However, in the intermediate quartiles (Q2 and Q3) no differences are observed between access modalities in relation to the uncited rate. Within the group of journals in intermediate positions in the rankings by impact factor, there is no observed OA citation advantage, at least in relation to a possible reduction of the uncited rate. This seems to indicate that access modality is not a determining factor for reading a paper, at least in the group of journals with a medium perceived quality.

In the case of conference proceedings, in half of the cases uncitedness is greater than 60%, much higher than that obtained for journals. For book series, OA books generally have a higher uncited rate than paywalled ones. Therefore, the OA does not guarantee better visibility and impact in disciplines where book series are widespread. Furthermore, the uncited rates for book series are much higher than for journals and conference proceedings. However, in the area in which the use of books is most widespread as a means of communication for research results (Humanities), the maturation time of citations is much longer, and therefore the citation window of two years after publication may be too short a period to measure the real impact in the medium and long term.

Finally, after removing the field effect, a strong negative correlation is observed between the uncited rate and the average impact. The higher the percentile of the serial title, the lower the uncited rate. This correlation coefficient is close to 0.95 in the case of journals, 0.97 in the case of conference proceedings, and falls to 0.90 in the case of book series. In general, a greater dispersion in the points cloud, and therefore a worse fit, is observed among the serial titles in the highest percentiles (Q1). Moreover, there are no differences according to access modality or size of the serial title.

With respect to the quality of the database, Scopus is among the largest citation databases, with a wide global and regional coverage of scientific journals, conference proceedings, and books. A rigorous content selection has allowed Scopus to be used as a bibliometric data source for large-scale analyses in research assessments, research landscape studies, science policy evaluations, and university rankings (Baas et al., 2020).

In relation to possible biases in the dataset, the use of either Elsevier's Scopus and Thomson Reuters' Web of Science (WoS) for research evaluation may introduce biases that favor Natural Sciences and Engineering as well as Biomedical Research to the detriment of Social Sciences and Arts and Humanities (Mongeon and Paul-Hus, 2016). These considerations imply that our results should be used with caution.

Regarding possible applications, the uncited rate of a serial title (journals, conference proceedings and book series) can be used as a better measure of the frequency of low-impact documents within that title than an impact ranking that measures the average impact of all published documents in that title. This is especially relevant in the case of interdisciplinary journals (Hernández and Dorta-González, 2020).

**References**

Baas, J., Schotten, M., Plume, A., Côté, G., & Karimi, R. (2020). Scopus as a curated, high-quality bibliometric data source for academic research in quantitative science studies. Quantitative Science Studies, 1(1), 377–386.

Burrell, Q. L. (2013). A stochastic approach to the relation between the impact factor and the uncitedness factor. Journal of Informetrics, 7(3), 676–682.

Dorta-González, P., González-Betancor, S. M., & Dorta-González, M.I. (2017). Reconsidering the gold open access citation advantage postulate in a multidisciplinary context: an analysis of the subject categories in the Web of Science database 2009-2014. Scientometrics, 112(2), 877–901.

Dorta-González, P., & Santana-Jiménez, Y. (2018). Prevalence and citation advantage of gold open access in the subject areas of the Scopus database. Research Evaluation, 27(1), 1–15.

Egghe, L. (2013). The functional relation between the impact factor and the uncitedness factor revisited. Journal of Informetrics, 7(1), 183–189.

Egghe, L., Guns, R., & Rousseau, R. (2011). Thoughts on uncitedness: Nobel laureates and fields medalists as case studies. Journal of the American Society for Information Science and Technology, 62(8), 1637–1644.

González-Betancor, S. M., & Dorta-González, P. (2019). Publication modalities 'article in press' and 'open access' in relation to journal average citation. Scientometrics, 120(3), 1209–1223.

Heneberg, P. (2013). Supposedly uncited articles of Nobel laureates and Fields medalists can be prevalently attributed to the errors of omission and commission. Journal of the American Society for Information Science and Technology, 64(3), 448–454.

Hernández, J. M., & Dorta-González, P. (2020). Interdisciplinarity metric based on the co-citation network. Mathematics, 8(4), 544.

Ho, Y. S., & Hartley, J. (2017). Sleeping beauties in psychology. Scientometrics, 110(1), 301–305.

Hsu, J. W., & Huang, D. W. (2012). A scaling between impact factor and uncitedness. Physica A: Statistical Mechanics and its Applications, 391(5), 2129–2134.

Hu, Z., & Wu, Y. (2014). Regularity in the time-dependent distribution of the percentage of never-cited papers: An empirical pilot study based on the six journals. Journal of Informetrics, 8(1), 136–146.

Kamat, P. V. (2018). Most cited versus uncited papers. What do they tell us? ACS Energy Letters, 3(9), 2134–2135.

Liang, L., Zhong, Z., & Rousseau, R. (2015). Uncited papers, uncited authors and uncited topics: A case study in library and information science. Journal of Informetrics, 9(1), 50–58.

Lou, W., & He, J. (2015). Does author affiliation reputation affect uncitedness? Proceedings of the Association for Information Science and Technology, 52(1), 1–4.

Mavrogenis, A. F., Quaile, A., Pećina, M., & Scarlat, M. M. (2018). Citations, non-citations and visibility of International Orthopaedics in 2017. International Orthopaedics, 42(11), 2499–2505.

Mongeon, P., & Paul-Hus, A. (2016). The journal coverage of Web of Science and Scopus: a comparative analysis. Scientometrics, 106(1), 213–228.

Nicolaisen, J., & Frandsen, T. F. (2019). Zero Impact: a large-scale study of uncitedness. Scientometrics, 119 (2), 1227–1254.

Rosenkrantz, A. B., Chung, R., & Duszak, R. (2019). Uncited research articles in popular United States general radiology journals. Academic Radiology, 26(2), 282–285

Thelwall, M. (2016). Are there too many uncited articles? Zero inflated variants of the discretised lognormal and hooked power law distributions. Journal of Informetrics, 10(2), 622–633.

Van Leeuwen, T. N., & Moed, H. F. (2005). Characteristics of journal impact factors: The effects of uncitedness and citation distribution on the understanding of journal impact factors. Scientometrics, 63(2), 357–371.

Van Noorden, R. (2017). The science that's never been cited. Nature, 552, 162–164.

Van Raan, A. F. (2015). Dormitory of physical and engineering sciences: Sleeping beauties may be sleeping innovations. PLoS ONE, 10(10), e0139786.

Wallace, M.L., Larivière, V., & Gingras, Y. (2009). Modeling a century of citation distributions. Journal of Informetrics, 3(4), 296–303.

Yamashita, Y., & Yoshinaga, D. (2014). Influence of researchers' international mobilities on publication: A comparison of highly cited and uncited papers. Scientometrics, 101(2), 1475–1489.

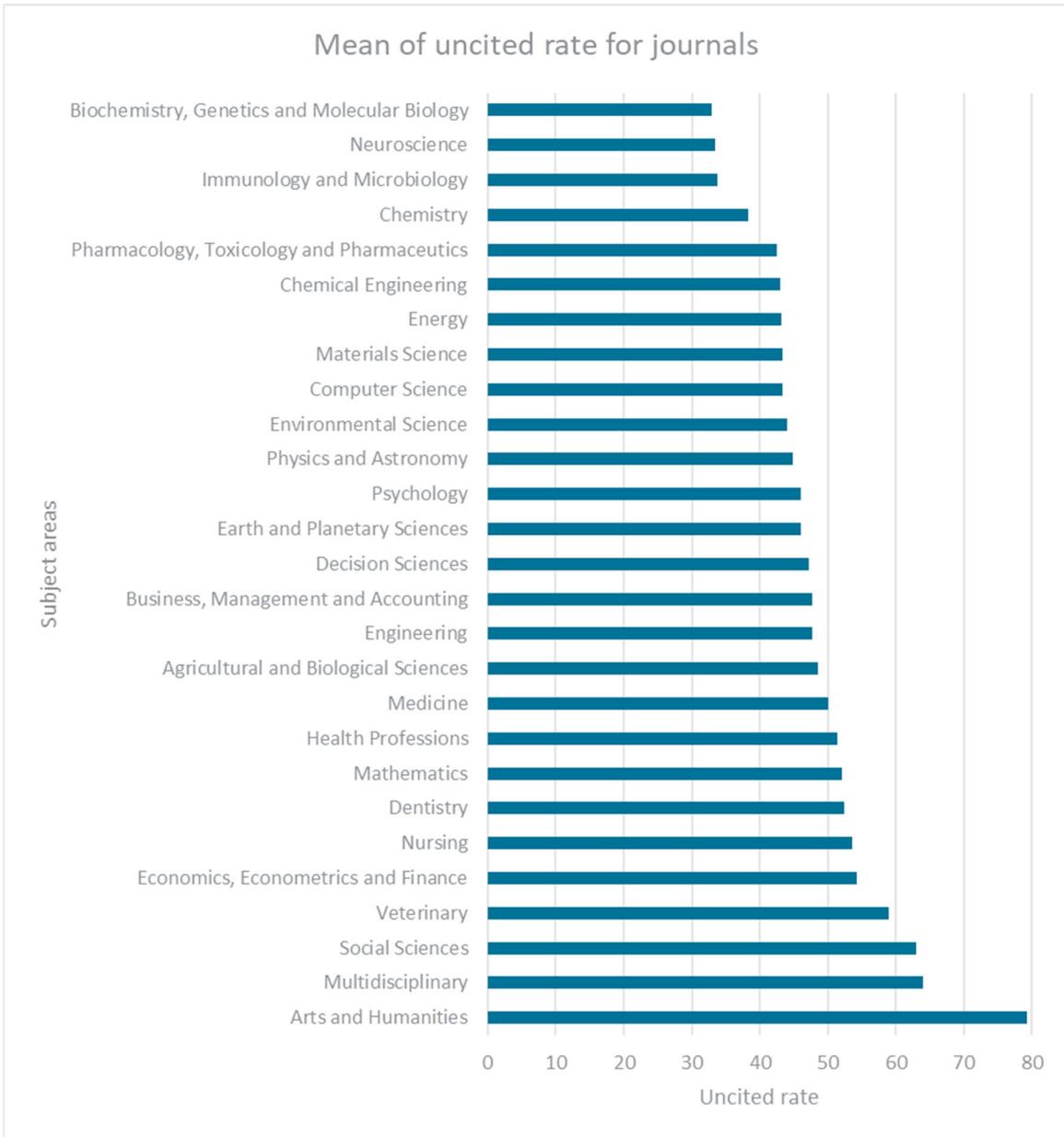

Figure 1: Mean of uncited rate for journals by subject area

Table 1: Central tendency and variability measures for the average uncited rate by access modality

|  | OA Journals | Paywalled Journals | All Journals |
|---|---|---|---|
| Median | 51 | 46 | 47 |
| Mean | 51.32 | 48.57 | 48.96 |
| Standard Deviation | 13.69 | 12.22 | 11.87 |
| Minimum | 18 | 23 | 23 |
| Maximum | 93 | 89 | 89 |
| Range of variation | 75 | 66 | 66 |
| Asymmetry coefficient | 0.544 | 0.840 | 0.835 |
| Kurtosis | 0.593 | 0.879 | 1.049 |

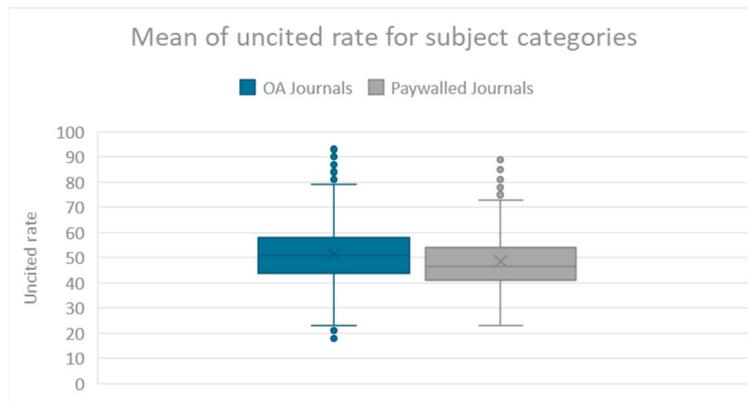

Figure 2: Distribution of mean uncited rate for subject categories by access modality

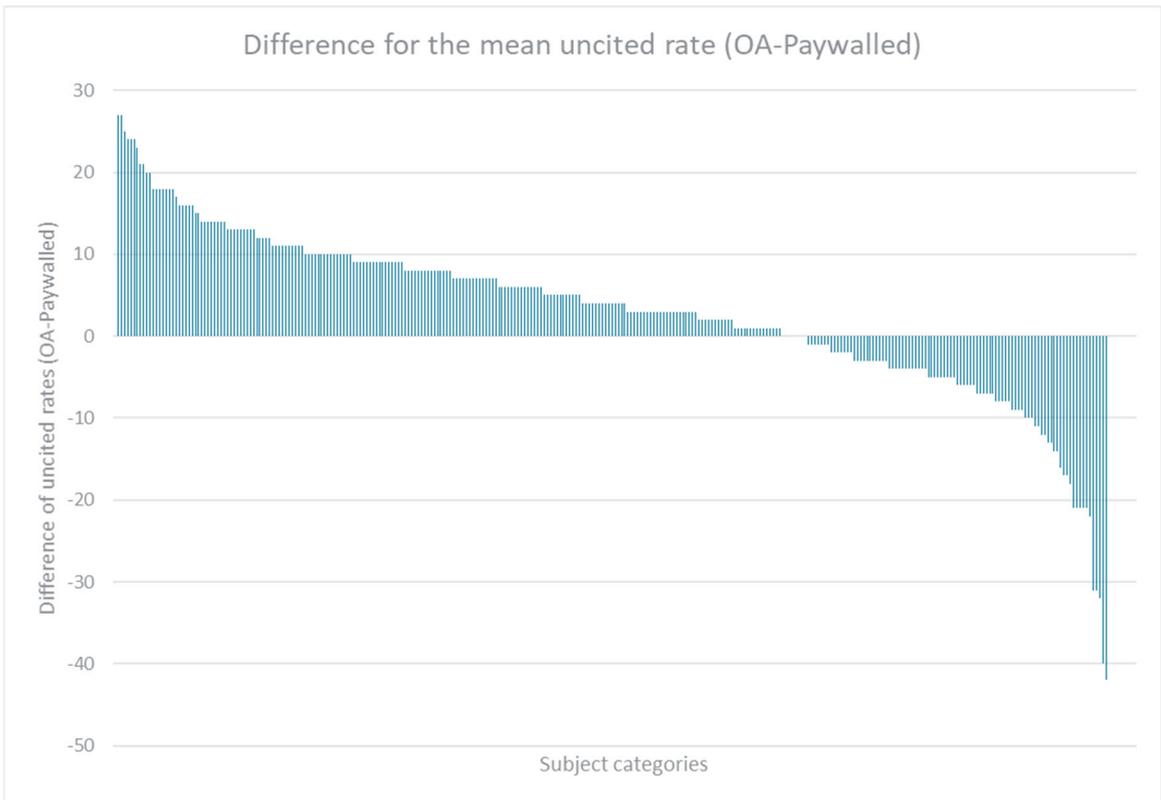

Figure 3: Difference in the mean uncited rate (OA - Paywalled) in journals by subject category

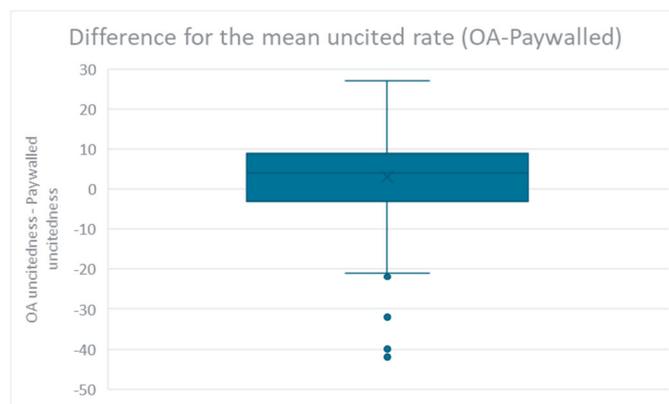

Figure 4: Distribution for the difference of mean uncited rates for journals by subject category

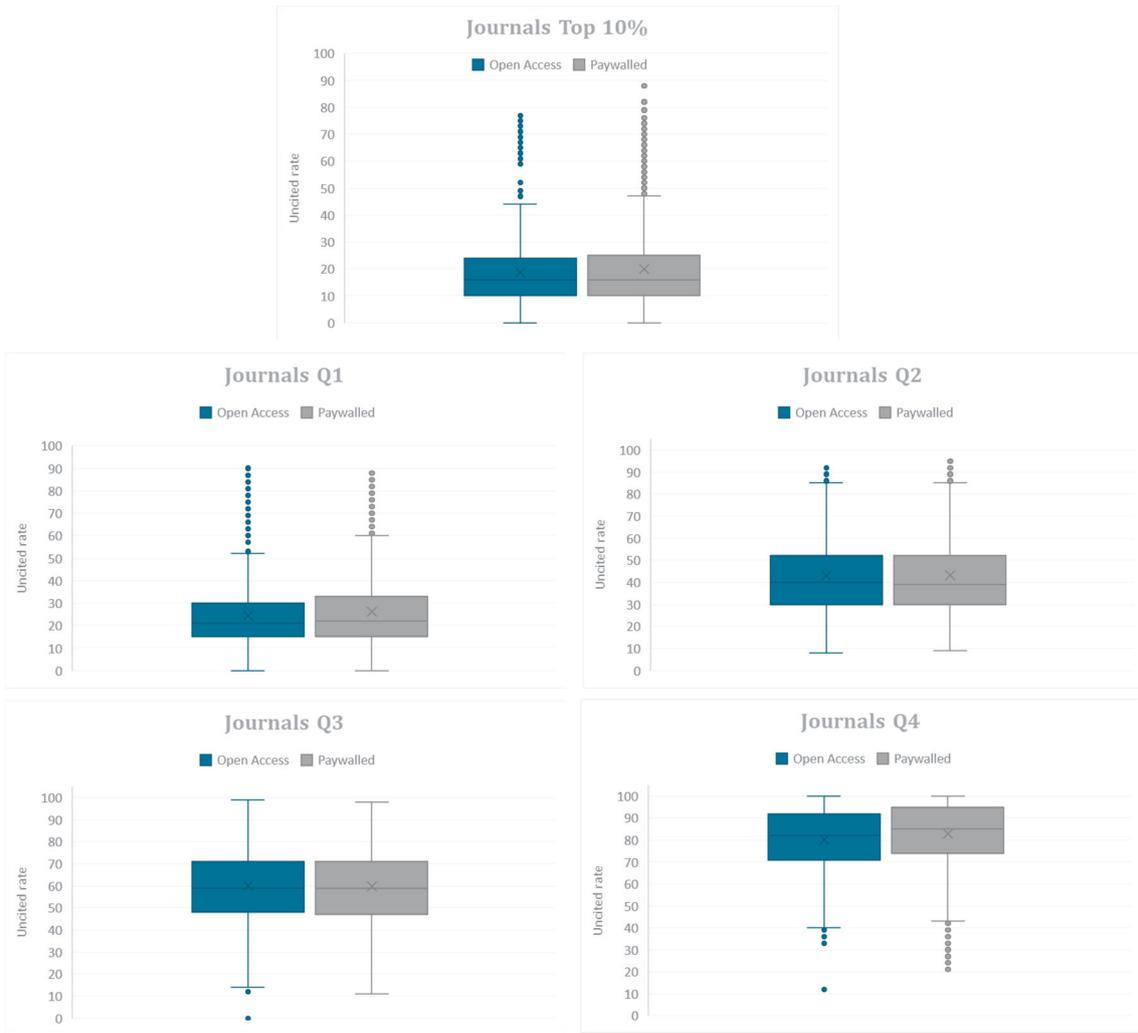

Figure 5: Box diagram for the distribution of the uncited rate for journals by access modality and average impact ranking (CiteScore)

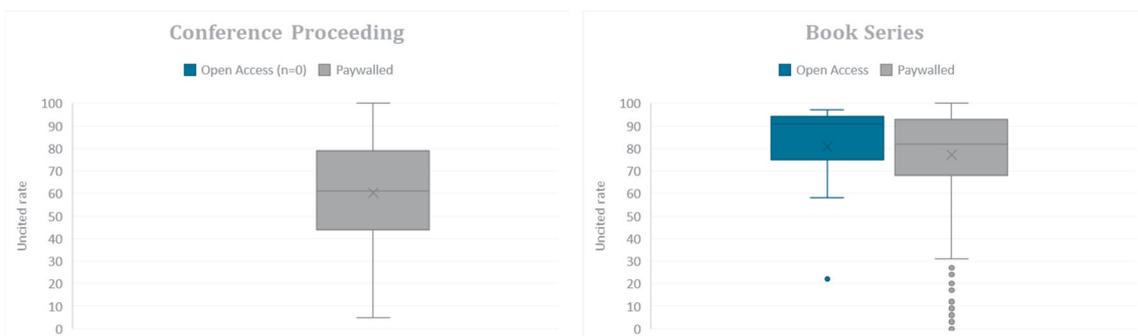

Figure 6: Box diagram for the distribution of the uncited rate for conference proceedings and book series by access modality

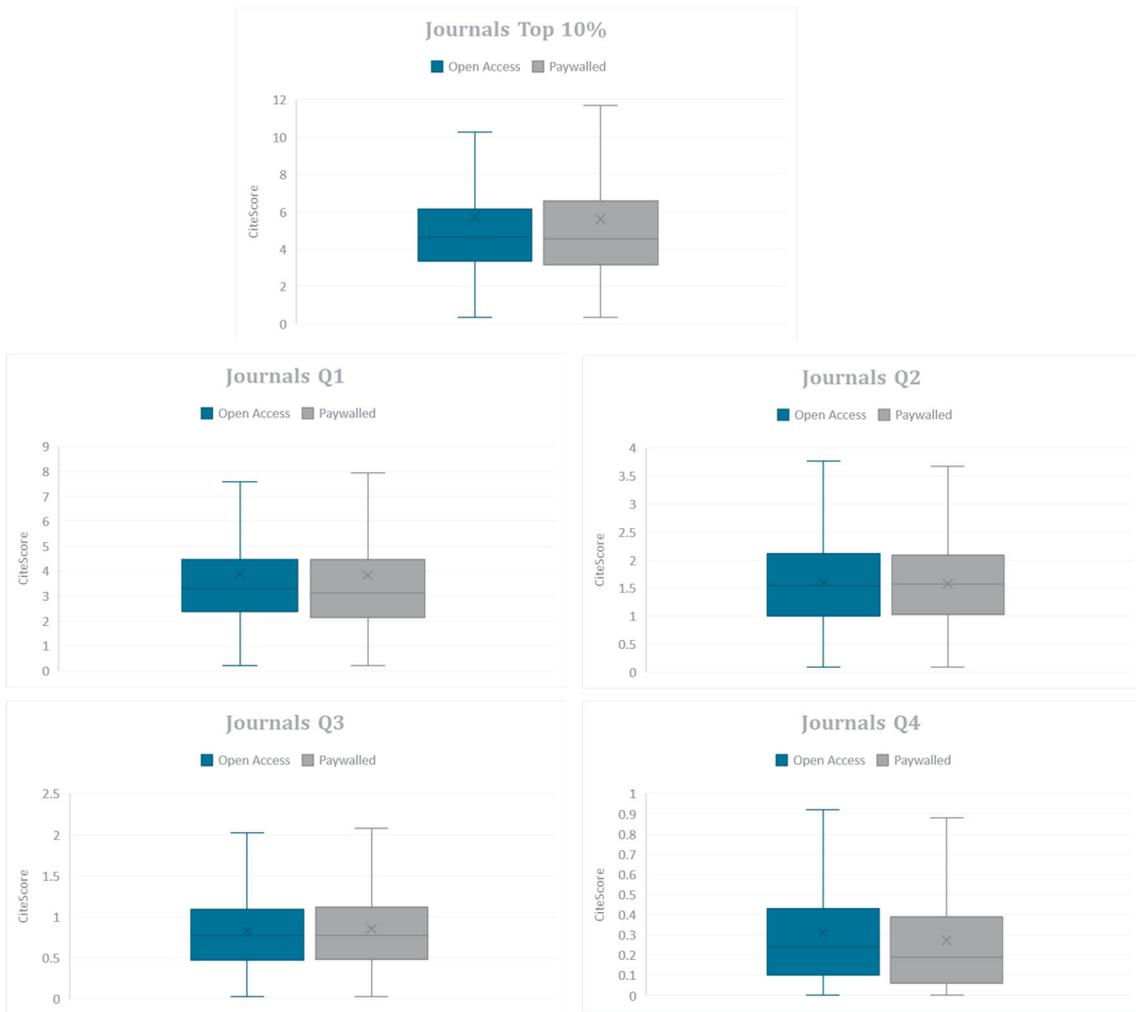

Figure 7: Box diagram for the distribution of CiteScore for journals by access modality and average impact ranking (outliers are not shown to allow better visualization of the vast majority of cases)

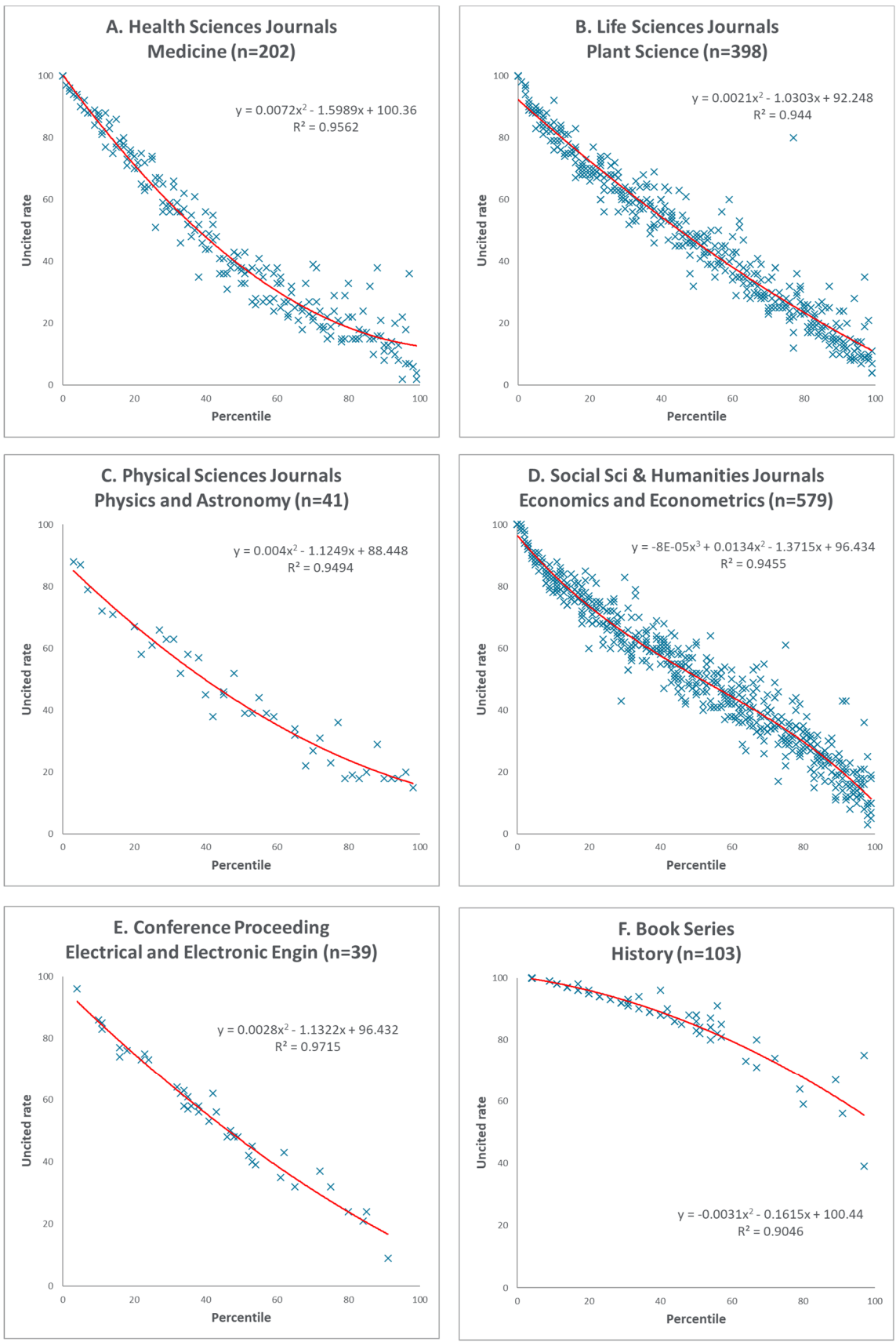

Figure 8: Scatter plot between the uncited rate and percentile in six subject categories

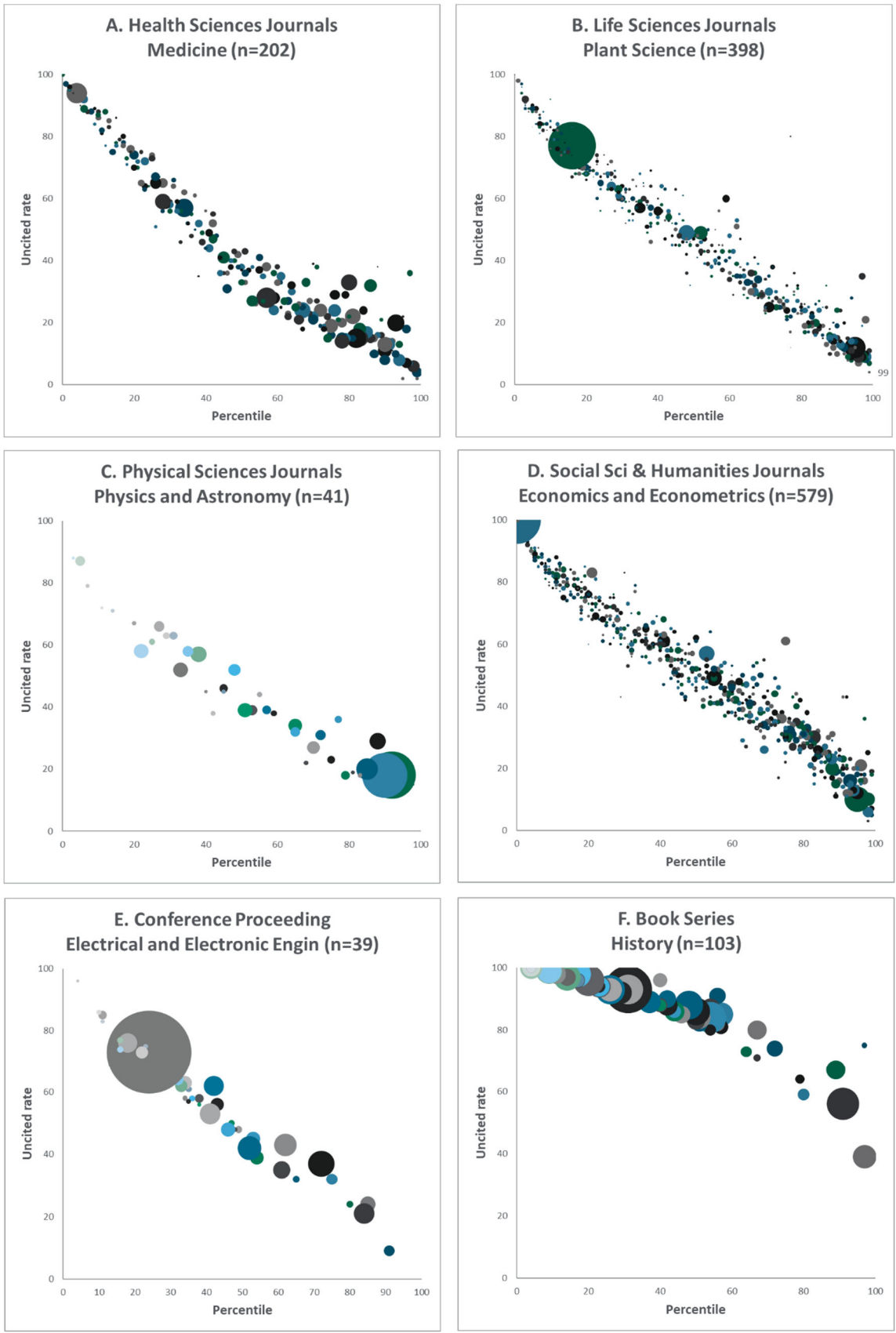

Figure 9: Bubble diagram for the source size (total documents) in six subject categories. The size of the bubble is proportional to the number of published documents in the serial title, and the coordinates are the uncited rate and percentile

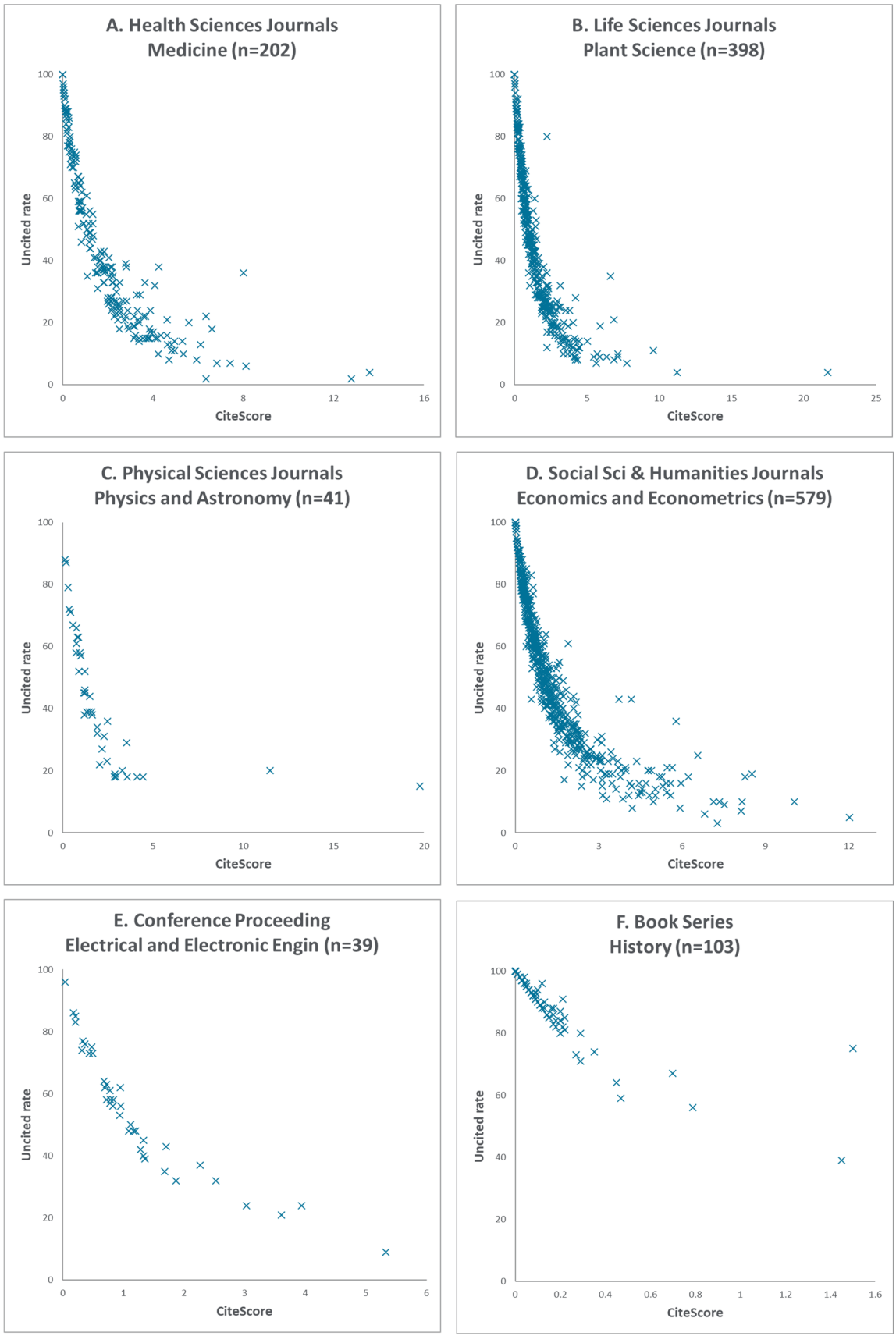

Figure 10: Scatter plot between uncited rate and CiteScore in six subject categories

**Appendix A:** Mean uncited rate for OA and paywalled journals by subject area and category

| Scopus Subject Area | Scopus Subject Category | Mean of Uncited Rate for Journals | | | | | | |
|---|---|---|---|---|---|---|---|---|
| | | Total Group | N | OA Group | N | Paywalled Group | N | Difference OA-Paywalled |
| Agricultural and Biological Sciences | General Agricultural and Biological Sciences | 56 | 177 | 55 | 79 | 57 | 98 | -2 |
| | Agricultural and Biological Sciences (miscellaneous) | 51 | 60 | 52 | 22 | 50 | 38 | 2 |
| | Agronomy and Crop Science | 51 | 317 | 55 | 105 | 48 | 212 | 7 |
| | Animal Science and Zoology | 52 | 380 | 57 | 133 | 50 | 247 | 7 |
| | Aquatic Science | 43 | 197 | 48 | 48 | 42 | 149 | 6 |
| | Ecology, Evolution, Behavior and Systematics | 45 | 577 | 48 | 143 | 44 | 434 | 4 |
| | Food Science | 46 | 266 | 48 | 77 | 46 | 189 | 2 |
| | Forestry | 52 | 132 | 53 | 50 | 52 | 82 | 1 |
| | Horticulture | 52 | 74 | 59 | 26 | 49 | 48 | 10 |
| | Insect Science | 51 | 138 | 55 | 36 | 50 | 102 | 5 |
| | Plant Science | 48 | 398 | 53 | 128 | 46 | 270 | 7 |
| | Soil Science | 45 | 114 | 55 | 39 | 40 | 75 | 15 |
| Arts and Humanities | General Arts and Humanities | 84 | 123 | 87 | 34 | 83 | 89 | 4 |
| | Arts and Humanities (miscellaneous) | 52 | 256 | 77 | 20 | 50 | 236 | 27 |
| | History | 81 | 1018 | 87 | 154 | 80 | 864 | 7 |
| | Language and Linguistics | 74 | 656 | 84 | 161 | 71 | 495 | 13 |
| | Archaeology | 73 | 473 | 78 | 90 | 72 | 383 | 6 |
| | Classics | 87 | 90 | 93 | 13 | 86 | 77 | 7 |
| | Conservation | 78 | 63 | 75 | 19 | 80 | 44 | -5 |
| | History and Philosophy of Science | 68 | 127 | 78 | 18 | 66 | 109 | 12 |
| | Literature and Literary Theory | 89 | 702 | 90 | 101 | 89 | 601 | 1 |
| | Museology | 80 | 42 | 82 | 9 | 80 | 33 | 2 |
| | Music | 83 | 120 | 91 | 14 | 82 | 106 | 9 |
| | Philosophy | 77 | 511 | 86 | 96 | 75 | 415 | 11 |
| | Religious studies | 85 | 398 | 88 | 45 | 85 | 353 | 3 |
| | Visual Arts and Performing Arts | 87 | 417 | 92 | 46 | 87 | 371 | 5 |
| Biochemistry, Genetics and Molecular Biology | General Biochemistry, Genetics and Molecular Biology | 44 | 183 | 44 | 79 | 44 | 104 | 0 |
| | Biochemistry, Genetics and Molecular Biology (miscellaneous) | 39 | 20 | 28 | 10 | 50 | 10 | -22 |
| | Ageing | 29 | 29 | 25 | 7 | 30 | 22 | -5 |
| | Biochemistry | 32 | 392 | 36 | 100 | 30 | 292 | 6 |
| | Biophysics | 38 | 122 | 47 | 25 | 36 | 97 | 11 |
| | Biotechnology | 40 | 246 | 42 | 75 | 39 | 171 | 3 |
| | Cancer Research | 31 | 187 | 26 | 43 | 32 | 144 | -6 |
| | Cell Biology | 28 | 263 | 26 | 62 | 29 | 201 | -3 |
| | Clinical Biochemistry | 35 | 112 | 31 | 25 | 35 | 87 | -4 |
| | Developmental Biology | 28 | 76 | 26 | 19 | 28 | 57 | -2 |
| | Endocrinology | 33 | 117 | 45 | 28 | 29 | 89 | 16 |
| | Genetics | 32 | 315 | 30 | 87 | 33 | 228 | -3 |
| | Molecular Biology | 29 | 371 | 29 | 102 | 29 | 269 | 0 |
| | Molecular Medicine | 30 | 159 | 28 | 42 | 31 | 117 | -3 |
| | Physiology | 32 | 164 | 35 | 28 | 31 | 136 | 4 |
| | Structural Biology | 33 | 46 | 27 | 10 | 35 | 36 | -8 |
| Business, Management and Accounting | General Business, Management and Accounting | 52 | 169 | 58 | 24 | 51 | 145 | 7 |
| | Business, Management and Accounting (miscellaneous) | 48 | 71 | 52 | 6 | 48 | 65 | 4 |
| | Accounting | 46 | 131 | 62 | 12 | 45 | 119 | 17 |
| | Business and International Management | 51 | 323 | 57 | 39 | 50 | 284 | 7 |
| | Management Information Systems | 44 | 76 | 53 | 5 | 43 | 71 | 10 |
| | Management of Technology and Innovation | 45 | 187 | 57 | 18 | 44 | 169 | 13 |
| | Marketing | 43 | 147 | 59 | 16 | 41 | 131 | 18 |
| | Organizational Behavior and Human Resource Management | 47 | 180 | 58 | 23 | 46 | 157 | 12 |
| | Strategy and Management | 48 | 383 | 53 | 46 | 47 | 337 | 6 |
| | Tourism, Leisure and Hospitality Management | 43 | 99 | 64 | 14 | 40 | 85 | 24 |
| | Industrial relations | 54 | 49 | 59 | 8 | 54 | 41 | 5 |
| Chemical Engineering | General Chemical Engineering | 44 | 257 | 47 | 50 | 44 | 207 | 3 |
| | Bioengineering | 35 | 131 | 39 | 26 | 33 | 105 | 6 |
| | Catalysis | 24 | 48 | 32 | 6 | 23 | 42 | 9 |
| | Chemical Health and Safety | 28 | 21 | 32 | 6 | 26 | 15 | 6 |
| | Filtration and Separation | 31 | 10 | 36 | 4 | 28 | 6 | 8 |
| | Fluid Flow and Transfer Processes | 45 | 123 | 45 | 29 | 44 | 94 | 1 |
| Chemistry | General Chemistry | 43 | 332 | 51 | 63 | 41 | 269 | 10 |
| | Chemistry (miscellaneous) | 37 | 23 | 34 | 6 | 39 | 17 | -5 |
| | Analytical Chemistry | 34 | 100 | 45 | 17 | 31 | 83 | 14 |
| | Electrochemistry | 34 | 30 | 44 | 5 | 32 | 25 | 12 |
| | Inorganic Chemistry | 37 | 64 | 45 | 7 | 36 | 57 | 9 |
| | Organic Chemistry | 34 | 159 | 44 | 21 | 33 | 138 | 11 |
| | Spectroscopy | 35 | 62 | 45 | 8 | 33 | 54 | 12 |
| Computer Science | General Computer Science | 50 | 180 | 52 | 54 | 49 | 126 | 3 |
| | Computer Science (miscellaneous) | 51 | 45 | 46 | 14 | 53 | 31 | -7 |
| | Artificial Intelligence | 40 | 171 | 48 | 29 | 39 | 142 | 9 |
| | Computational Theory and Mathematics | 45 | 110 | 53 | 16 | 44 | 94 | 9 |
| | Computer Graphics and Computer-Aided Design | 43 | 65 | 44 | 10 | 43 | 55 | 1 |
| | Computer Networks and Communications | 46 | 246 | 48 | 51 | 45 | 195 | 3 |
| | Computer Science Applications | 43 | 528 | 46 | 103 | 42 | 425 | 4 |
| | Computer Vision and Pattern Recognition | 42 | 70 | 51 | 11 | 40 | 59 | 11 |
| | Hardware and Architecture | 42 | 132 | 45 | 17 | 42 | 115 | 3 |
| | Human-Computer Interaction | 40 | 86 | 55 | 14 | 37 | 72 | 18 |
| | Information Systems | 44 | 253 | 52 | 44 | 42 | 209 | 10 |
| | Signal Processing | 40 | 87 | 51 | 22 | 37 | 65 | 14 |
| | Software | 41 | 311 | 54 | 30 | 39 | 281 | 15 |
| Decision Sciences | General Decision Sciences | 38 | 28 | 46 | 3 | 37 | 25 | 9 |
| | Decision Sciences (miscellaneous) | 46 | 6 | | 0 | 46 | 6 | -46 |
| | Information Systems and Management | 47 | 88 | 56 | 12 | 46 | 76 | 10 |
| | Management Science and Operations Research | 45 | 145 | 53 | 18 | 44 | 127 | 9 |
| | Statistics, Probability and Uncertainty | 52 | 120 | 54 | 19 | 51 | 101 | 3 |
| Dentistry | General Dentistry | 53 | 110 | 58 | 44 | 50 | 66 | 8 |
| | Dentistry (miscellaneous) | 52 | 15 | 60 | 8 | 42 | 7 | 18 |
| | Oral Surgery | 53 | 46 | 70 | 12 | 47 | 34 | 23 |
| | Orthodontics | 60 | 18 | 58 | 6 | 61 | 12 | -3 |
| | Periodontics | 42 | 21 | 63 | 5 | 36 | 16 | 27 |
| Earth and Planetary Sciences | General Earth and Planetary Sciences | 53 | 170 | 49 | 60 | 55 | 110 | -6 |
| | Earth and Planetary Sciences (miscellaneous) | 46 | 84 | 47 | 22 | 46 | 62 | 1 |
| | Atmospheric Science | 38 | 106 | 40 | 34 | 37 | 72 | 3 |
| | Computers in Earth Sciences | 41 | 30 | 46 | 10 | 38 | 20 | 8 |
| | Earth-Surface Processes | 51 | 127 | 53 | 33 | 51 | 94 | 2 |
| | Economic Geology | 47 | 30 | 57 | 7 | 44 | 23 | 13 |
| | Geochemistry and Petrology | 38 | 111 | 41 | 19 | 38 | 92 | 3 |
| | Geology | 48 | 205 | 54 | 57 | 46 | 148 | 8 |
| | Geophysics | 44 | 98 | 50 | 23 | 42 | 75 | 8 |
| | Geotechnical Engineering and Engineering Geology | 48 | 154 | 48 | 30 | 47 | 124 | 1 |
| | Oceanography | 45 | 111 | 47 | 27 | 44 | 84 | 3 |
| | Paleontology | 49 | 88 | 52 | 26 | 47 | 62 | 5 |
| | Space and Planetary Science | 39 | 79 | 38 | 11 | 40 | 68 | -2 |
| | Stratigraphy | 46 | 34 | 49 | 14 | 44 | 20 | 5 |
| Economics, Econometrics and Finance | General Economics, Econometrics and Finance | 63 | 196 | 66 | 52 | 62 | 144 | 4 |
| | Economics, Econometrics and Finance (miscellaneous) | 56 | 96 | 60 | 23 | 55 | 73 | 5 |
| | Economics and Econometrics | 52 | 579 | 61 | 62 | 51 | 517 | 10 |
| | Finance | 52 | 244 | 62 | 29 | 51 | 215 | 11 |
| Energy | General Energy | 48 | 61 | 43 | 11 | 48 | 50 | -5 |
| | Energy (miscellaneous) | 43 | 19 | 50 | 6 | 39 | 13 | 11 |
| | Energy Engineering and Power Technology | 44 | 163 | 40 | 28 | 44 | 135 | -4 |
| | Fuel Technology | 41 | 81 | 36 | 15 | 43 | 66 | -7 |
| | Nuclear Energy and Engineering | 53 | 57 | 51 | 9 | 54 | 48 | -3 |
| | Renewable Energy, Sustainability and the Environment | 38 | 148 | 37 | 34 | 38 | 114 | -1 |
| Engineering | General Engineering | 55 | 252 | 53 | 65 | 56 | 187 | -3 |
| | Engineering (miscellaneous) | 52 | 51 | 47 | 12 | 53 | 39 | -6 |
| | Aerospace Engineering | 48 | 107 | 54 | 19 | 47 | 88 | 7 |
| | Automotive Engineering | 52 | 76 | 59 | 22 | 49 | 54 | 10 |
| | Biomedical Engineering | 38 | 196 | 35 | 46 | 39 | 150 | -4 |
| | Civil and Structural Engineering | 45 | 260 | 50 | 57 | 45 | 203 | 5 |
| | Computational Mechanics | 47 | 56 | 54 | 15 | 45 | 41 | 9 |
| | Control and Systems Engineering | 42 | 212 | 52 | 34 | 40 | 178 | 12 |
| | Electrical and Electronic Engineering | 47 | 569 | 52 | 84 | 46 | 485 | 6 |
| | Industrial and Manufacturing Engineering | 47 | 769 | 50 | 125 | 46 | 644 | 4 |
| | Mechanics of Materials | 44 | 327 | 51 | 48 | 43 | 279 | 8 |
| | Ocean Engineering | 51 | 80 | 47 | 16 | 52 | 64 | -5 |
| | Safety, Risk, Reliability and Quality | 50 | 139 | 54 | 18 | 50 | 121 | 4 |
| | Media Technology | 61 | 46 | 65 | 5 | 60 | 41 | 5 |
| | Building and Construction | 47 | 144 | 58 | 25 | 45 | 119 | 13 |
| | Architecture | 73 | 104 | 74 | 27 | 73 | 77 | 1 |
| Environmental Science | General Environmental Science | 48 | 180 | 56 | 41 | 46 | 139 | 10 |
| | Environmental Science (miscellaneous) | 45 | 68 | 48 | 19 | 44 | 49 | 4 |
| | Ecological Modelling | 36 | 29 | 33 | 9 | 37 | 20 | -4 |
| | Ecology | 47 | 325 | 52 | 100 | 45 | 225 | 7 |
| | Environmental Chemistry | 30 | 95 | 52 | 8 | 28 | 87 | 24 |
| | Environmental Engineering | 42 | 114 | 47 | 24 | 41 | 90 | 6 |
| | Global and Planetary Change | 35 | 67 | 37 | 18 | 34 | 49 | 3 |
| | Health, Toxicology and Mutagenesis | 37 | 113 | 35 | 30 | 38 | 83 | -3 |
| | Management, Monitoring, Policy and Law | 45 | 273 | 46 | 59 | 45 | 214 | 1 |
| | Nature and Landscape Conservation | 48 | 136 | 46 | 46 | 49 | 90 | -3 |
| | Pollution | 42 | 105 | 47 | 16 | 41 | 89 | 6 |
| | Waste Management and Disposal | 44 | 87 | 48 | 21 | 43 | 66 | 5 |

| Category | Subcategory | | | | | | | |
|---|---|---|---|---|---|---|---|---|
| | Water Science and Technology | 48 | 192 | 43 | 43 | 49 | 149 | -6 |
| Health Professions | Health Professions (miscellaneous) | 56 | 16 | 65 | 2 | 55 | 14 | 10 |
| | Chiropractics | 54 | 5 | | 0 | 54 | 5 | -54 |
| | Complementary and Manual Therapy | 64 | 13 | 50 | 1 | 66 | 12 | -16 |
| | Health Information Management | 44 | 24 | 39 | 8 | 47 | 16 | -8 |
| | Medical Laboratory Technology | 53 | 27 | 51 | 5 | 54 | 22 | -3 |
| | Occupational Therapy | 54 | 15 | 58 | 2 | 54 | 13 | 4 |
| | Optometry | 54 | 8 | 43 | 1 | 55 | 7 | -12 |
| | Pharmacy | 65 | 23 | 57 | 6 | 68 | 17 | -11 |
| | Physical Therapy, Sports Therapy and Rehabilitation | 51 | 176 | 53 | 49 | 51 | 127 | 2 |
| | Podiatry | 64 | 6 | 31 | 1 | 71 | 5 | -40 |
| | Radiological and Ultrasound Technology | 43 | 46 | 44 | 13 | 43 | 33 | 1 |
| | Speech and Hearing | 49 | 53 | 55 | 6 | 49 | 47 | 6 |
| Immunology and Microbiology | General Immunology and Microbiology | 34 | 41 | 35 | 14 | 33 | 27 | 2 |
| | Immunology and Microbiology (miscellaneous) | 43 | 5 | 18 | 2 | 60 | 3 | -42 |
| | Applied Microbiology and Biotechnology | 38 | 98 | 35 | 22 | 39 | 76 | -4 |
| | Immunology | 31 | 197 | 32 | 52 | 31 | 145 | 1 |
| | Microbiology | 32 | 138 | 34 | 40 | 31 | 98 | 3 |
| | Parasitology | 37 | 61 | 34 | 29 | 39 | 32 | -5 |
| | Virology | 36 | 65 | 26 | 19 | 40 | 46 | -14 |
| Materials Science | General Materials Science | 43 | 388 | 47 | 65 | 42 | 323 | 5 |
| | Materials Science (miscellaneous) | 52 | 69 | 46 | 24 | 55 | 45 | -9 |
| | Biomaterials | 30 | 80 | 33 | 17 | 29 | 63 | 4 |
| | Ceramics and Composites | 42 | 94 | 42 | 17 | 42 | 77 | 0 |
| | Electronic, Optical and Magnetic Materials | 42 | 204 | 44 | 38 | 41 | 166 | 3 |
| | Materials Chemistry | 44 | 248 | 50 | 23 | 43 | 225 | 7 |
| | Metals and Alloys | 52 | 134 | 49 | 23 | 53 | 111 | -4 |
| | Polymers and Plastics | 42 | 126 | 48 | 12 | 42 | 114 | 6 |
| | Surfaces, Coatings and Films | 43 | 105 | 43 | 18 | 43 | 87 | 0 |
| Mathematics | General Mathematics | 61 | 319 | 68 | 64 | 59 | 255 | 9 |
| | Mathematics (miscellaneous) | 59 | 38 | 71 | 8 | 55 | 30 | 16 |
| | Algebra and Number Theory | 59 | 86 | 61 | 15 | 59 | 71 | 2 |
| | Analysis | 55 | 131 | 62 | 28 | 53 | 103 | 9 |
| | Applied Mathematics | 50 | 439 | 58 | 70 | 48 | 369 | 10 |
| | Computational Mathematics | 47 | 128 | 49 | 21 | 47 | 107 | 2 |
| | Control and Optimization | 47 | 83 | 49 | 18 | 46 | 65 | 3 |
| | Discrete Mathematics and Combinatorics | 58 | 62 | 65 | 15 | 55 | 47 | 10 |
| | Geometry and Topology | 57 | 75 | 59 | 17 | 56 | 58 | 3 |
| | Logic | 57 | 26 | 79 | 3 | 54 | 23 | 25 |
| | Mathematical Physics | 50 | 54 | 61 | 8 | 48 | 46 | 13 |
| | Modelling and Simulation | 44 | 241 | 46 | 43 | 43 | 198 | 3 |
| | Numerical Analysis | 49 | 49 | 66 | 7 | 46 | 42 | 20 |
| | Statistics and Probability | 53 | 200 | 55 | 34 | 52 | 166 | 3 |
| | Theoretical Computer Science | 44 | 108 | 55 | 15 | 42 | 93 | 13 |
| Medicine | General Medicine | 69 | 548 | 65 | 187 | 71 | 361 | -6 |
| | Medicine (miscellaneous) | 45 | 202 | 43 | 52 | 45 | 150 | -2 |
| | Anatomy | 48 | 36 | 61 | 6 | 45 | 30 | 16 |
| | Anesthesiology and Pain Medicine | 58 | 115 | 56 | 31 | 59 | 84 | -3 |
| | Biochemistry, medical | 39 | 49 | 36 | 18 | 41 | 31 | -5 |
| | Cardiology and Cardiovascular Medicine | 53 | 319 | 58 | 76 | 51 | 243 | 7 |
| | Critical Care and Intensive Care Medicine | 58 | 80 | 57 | 21 | 59 | 59 | -2 |
| | Complementary and alternative medicine | 57 | 83 | 41 | 19 | 62 | 64 | -21 |
| | Dermatology | 55 | 128 | 56 | 41 | 55 | 87 | 1 |
| | Embryology | 40 | 14 | | 0 | 40 | 14 | -40 |
| | Emergency Medicine | 63 | 77 | 57 | 18 | 66 | 59 | -9 |
| | Endocrinology, Diabetes and Metabolism | 42 | 205 | 45 | 69 | 41 | 136 | 4 |
| | Epidemiology | 36 | 90 | 31 | 33 | 39 | 57 | -8 |
| | Family Practice | 67 | 36 | 57 | 19 | 78 | 17 | -21 |
| | Gastroenterology | 51 | 131 | 46 | 38 | 53 | 93 | -7 |
| | Genetics (clinical) | 33 | 90 | 30 | 24 | 34 | 66 | -4 |
| | Geriatrics and Gerontology | 43 | 92 | 51 | 16 | 41 | 76 | 10 |
| | Health Informatics | 43 | 60 | 37 | 21 | 46 | 39 | -9 |
| | Health Policy | 52 | 222 | 46 | 52 | 54 | 170 | -8 |
| | Hematology | 45 | 119 | 44 | 28 | 45 | 91 | -1 |
| | Hepatology | 47 | 56 | 46 | 15 | 47 | 41 | -1 |
| | Histology | 43 | 58 | 37 | 16 | 44 | 42 | -7 |
| | Immunology and Allergy | 37 | 186 | 39 | 49 | 36 | 137 | 3 |
| | Internal Medicine | 50 | 119 | 47 | 28 | 51 | 91 | -4 |
| | Infectious Diseases | 44 | 271 | 41 | 105 | 46 | 166 | -5 |
| | Microbiology (medical) | 41 | 109 | 38 | 40 | 42 | 69 | -4 |
| | Nephrology | 47 | 56 | 52 | 20 | 44 | 36 | 8 |
| | Clinical Neurology | 45 | 332 | 47 | 80 | 45 | 252 | 2 |
| | Obstetrics and Gynecology | 54 | 165 | 55 | 39 | 54 | 126 | 1 |
| | Oncology | 41 | 317 | 36 | 90 | 43 | 227 | -7 |
| | Ophthalmology | 53 | 110 | 53 | 29 | 53 | 81 | 0 |
| | Orthopedics and Sports Medicine | 50 | 244 | 50 | 67 | 50 | 177 | 0 |
| | Otorhinolaryngology | 55 | 99 | 51 | 18 | 55 | 81 | -4 |
| | Pathology and Forensic Medicine | 48 | 182 | 47 | 28 | 49 | 154 | -2 |
| | Pediatrics, Perinatology, and Child Health | 56 | 273 | 63 | 54 | 54 | 219 | 9 |
| | Pharmacology (medical) | 47 | 230 | 37 | 50 | 50 | 180 | -13 |
| | Physiology (medical) | 39 | 94 | 44 | 17 | 38 | 77 | 6 |
| | Psychiatry and Mental health | 56 | 37 | 60 | 6 | 55 | 31 | 5 |
| | Public Health, Environmental and Occupational Health | 51 | 124 | 62 | 21 | 49 | 103 | 13 |
| | Pulmonary and Respiratory Medicine | 47 | 131 | 41 | 44 | 51 | 87 | -10 |
| | Radiology Nuclear Medicine and imaging | 47 | 267 | 46 | 75 | 48 | 192 | -2 |
| | Rehabilitation | 54 | 109 | 55 | 19 | 54 | 90 | 1 |
| | Reproductive Medicine | 47 | 63 | 47 | 18 | 48 | 45 | -1 |
| | Rheumatology | 41 | 53 | 44 | 18 | 40 | 35 | 4 |
| | Surgery | 56 | 389 | 61 | 94 | 54 | 295 | 7 |
| | Transplantation | 44 | 36 | 58 | 9 | 40 | 27 | 18 |
| | Urology | 53 | 99 | 55 | 33 | 53 | 66 | 2 |
| Multidisciplinary | Multidisciplinary | 64 | 89 | 52 | 29 | 69 | 60 | -17 |
| Neuroscience | General Neuroscience | 35 | 109 | 40 | 31 | 33 | 78 | 7 |
| | Neuroscience (miscellaneous) | 33 | 24 | 35 | 12 | 30 | 12 | 5 |
| | Behavioral Neuroscience | 32 | 68 | 32 | 17 | 32 | 51 | 0 |
| | Biological Psychiatry | 35 | 40 | 36 | 12 | 35 | 28 | 1 |
| | Cellular and Molecular Neuroscience | 23 | 85 | 21 | 29 | 25 | 56 | -4 |
| | Cognitive Neuroscience | 32 | 91 | 28 | 23 | 34 | 68 | -6 |
| | Developmental Neuroscience | 32 | 34 | 23 | 9 | 35 | 25 | -12 |
| | Endocrine and Autonomic Systems | 32 | 24 | 44 | 6 | 28 | 18 | 16 |
| | Neurology | 40 | 146 | 42 | 38 | 39 | 108 | 3 |
| | Sensory Systems | 35 | 39 | 35 | 8 | 35 | 31 | 0 |
| Nursing | General Nursing | 59 | 109 | 59 | 21 | 60 | 88 | -1 |
| | Nursing (miscellaneous) | 58 | 18 | 78 | 1 | 57 | 17 | 21 |
| | Advanced and Specialised Nursing | 68 | 53 | 52 | 3 | 69 | 50 | -17 |
| | Assessment and Diagnosis | 69 | 6 | | 0 | 69 | 6 | -69 |
| | Care Planning | 60 | 6 | | 0 | 60 | 6 | -60 |
| | Community and Home Care | 61 | 33 | 61 | 8 | 60 | 25 | 1 |
| | Critical Care | 62 | 18 | 69 | 1 | 61 | 17 | 8 |
| | Emergency | 67 | 23 | 58 | 3 | 68 | 20 | -10 |
| | Fundamentals and skills | 67 | 13 | 38 | 1 | 70 | 12 | -32 |
| | Gerontology | 51 | 33 | 61 | 3 | 50 | 30 | 11 |
| | Issues, ethics and legal aspects | 56 | 36 | 38 | 4 | 59 | 32 | -21 |
| | Leadership and Management | 62 | 28 | 45 | 2 | 63 | 26 | -18 |
| | LPN and LVN | 62 | 16 | | 0 | 62 | 16 | -62 |
| | Maternity and Midwifery | 65 | 23 | 37 | 2 | 68 | 21 | -31 |
| | Medical–Surgical | 72 | 23 | 59 | 1 | 72 | 22 | -13 |
| | Nutrition and Dietetics | 43 | 113 | 42 | 22 | 43 | 91 | -1 |
| | Oncology (nursing) | 47 | 15 | | 0 | 47 | 15 | -47 |
| | Pediatrics | 59 | 21 | 51 | 2 | 60 | 19 | -9 |
| | Pharmacology (nursing) | 81 | 6 | | 0 | 81 | 6 | -81 |
| | Psychiatric Mental Health | 36 | 145 | 44 | 17 | 35 | 128 | 9 |
| | Research and Theory | 54 | 9 | | 0 | 54 | 9 | -54 |
| Pharmacology, Toxicology and Pharmaceutics | General Pharmacology, Toxicology and Pharmaceutics | 51 | 59 | 44 | 29 | 58 | 30 | -14 |
| | Pharmacology, Toxicology and Pharmaceutics (miscellaneous) | 70 | 17 | 46 | 4 | 77 | 13 | -31 |
| | Drug Discovery | 38 | 139 | 39 | 30 | 37 | 109 | 2 |
| | Pharmaceutical Science | 49 | 152 | 47 | 39 | 50 | 113 | -3 |
| | Pharmacology | 41 | 293 | 44 | 60 | 41 | 233 | 3 |
| | Toxicology | 34 | 110 | 42 | 22 | 31 | 88 | 11 |
| Physics and Astronomy | General Physics and Astronomy | 48 | 206 | 54 | 43 | 46 | 163 | 8 |
| | Physics and Astronomy (miscellaneous) | 43 | 41 | 49 | 12 | 41 | 29 | 8 |
| | Acoustics and Ultrasonics | 47 | 39 | 38 | 7 | 48 | 32 | -10 |
| | Astronomy and Astrophysics | 46 | 72 | 57 | 11 | 44 | 61 | 13 |
| | Condensed Matter Physics | 43 | 375 | 50 | 32 | 42 | 343 | 8 |
| | Instrumentation | 45 | 113 | 37 | 25 | 48 | 88 | -11 |
| | Nuclear and High Energy Physics | 50 | 67 | 44 | 13 | 51 | 54 | -7 |
| | Atomic and Molecular Physics, and Optics | 42 | 162 | 39 | 27 | 43 | 135 | -4 |
| | Radiation | 49 | 48 | 45 | 10 | 50 | 38 | -5 |
| | Statistical and Nonlinear Physics | 48 | 42 | 28 | 3 | 49 | 39 | -21 |
| | Surfaces and Interfaces | 41 | 50 | 47 | 4 | 40 | 46 | 7 |
| Psychology | General Psychology | 49 | 192 | 63 | 45 | 45 | 147 | 18 |
| | Psychology (miscellaneous) | 47 | 36 | 59 | 3 | 45 | 33 | 14 |
| | Applied Psychology | 43 | 215 | 52 | 21 | 42 | 194 | 10 |
| | Clinical Psychology | 52 | 262 | 64 | 30 | 51 | 232 | 13 |
| | Developmental and Educational Psychology | 44 | 283 | 63 | 20 | 43 | 263 | 20 |
| | Experimental and Cognitive Psychology | 39 | 133 | 52 | 8 | 38 | 125 | 14 |
| | Neuropsychology and Physiological Psychology | 41 | 58 | 48 | 8 | 40 | 50 | 8 |
| | Social Psychology | 47 | 254 | 55 | 30 | 46 | 224 | 9 |
| Social Sciences | General Social Sciences | 65 | 213 | 77 | 53 | 61 | 160 | 16 |
| | Social Sciences (miscellaneous) | 55 | 235 | 73 | 35 | 52 | 200 | 21 |
| | Archaeology | 73 | 473 | 78 | 90 | 72 | 383 | 6 |
| | Development | 55 | 210 | 63 | 32 | 53 | 178 | 10 |
| | Education | 55 | 1021 | 62 | 186 | 53 | 835 | 9 |
| | Geography, Planning and Development | 55 | 608 | 62 | 120 | 54 | 488 | 8 |
| | Health (social science) | 51 | 244 | 54 | 48 | 51 | 196 | 3 |
| | Human Factors and Ergonomics | 43 | 34 | 60 | 2 | 42 | 32 | 18 |
| | Law | 65 | 574 | 75 | 79 | 64 | 495 | 11 |
| | Library and Information Sciences | 63 | 204 | 69 | 47 | 61 | 157 | 8 |
|  | Linguistics and Language | 73 | 703 | 84 | 168 | 70 | 535 | 14 |
|  | Safety Research | 51 | 67 | 45 | 18 | 53 | 49 | -8 |
|  | Sociology and Political Science | 61 | 1065 | 73 | 166 | 59 | 899 | 14 |
|  | Transportation | 42 | 89 | 56 | 19 | 38 | 70 | 18 |
|  | Anthropology | 67 | 328 | 76 | 66 | 65 | 262 | 11 |
|  | Communication | 61 | 307 | 73 | 48 | 59 | 259 | 14 |
|  | Cultural Studies | 78 | 829 | 81 | 137 | 78 | 692 | 3 |
|  | Demography | 60 | 93 | 67 | 22 | 57 | 71 | 10 |
|  | Gender Studies | 60 | 123 | 72 | 15 | 58 | 108 | 14 |
|  | Life-span and Life-course Studies | 43 | 46 | 65 | 5 | 41 | 41 | 24 |
|  | Political Science and International Relations | 63 | 458 | 71 | 64 | 62 | 394 | 9 |
|  | Urban Studies | 60 | 154 | 66 | 43 | 58 | 111 | 8 |
| Veterinary | General Veterinary | 61 | 166 | 64 | 71 | 60 | 95 | 4 |
|  | Veterinary (miscellaneous) | 50 | 9 | 53 | 4 | 47 | 5 | 6 |
|  | Equine | 49 | 7 | 48 | 1 | 49 | 6 | -1 |
|  | Food Animals | 51 | 29 | 58 | 4 | 49 | 25 | 9 |
|  | Small Animals | 62 | 17 | 42 | 1 | 63 | 16 | -21 |

**Appendix B**: Scatter plot for OA journals in four subject categories (there are no OA conference proceedings in Electrical and Electronic Engineering, and only two OA book series in History)

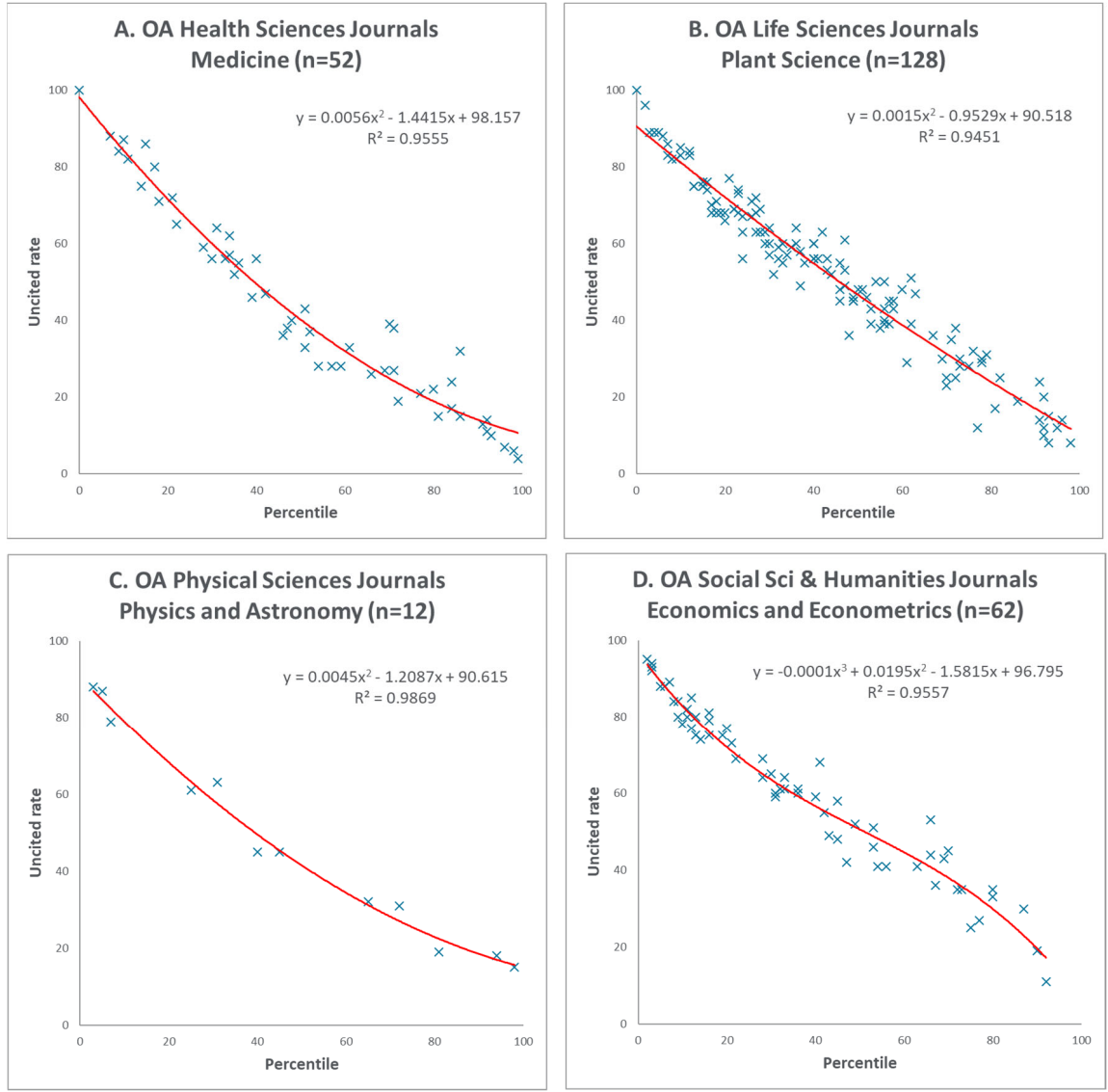

Figure B.1: Scatter plot between uncited rate and percentile for OA journals in four subject categories

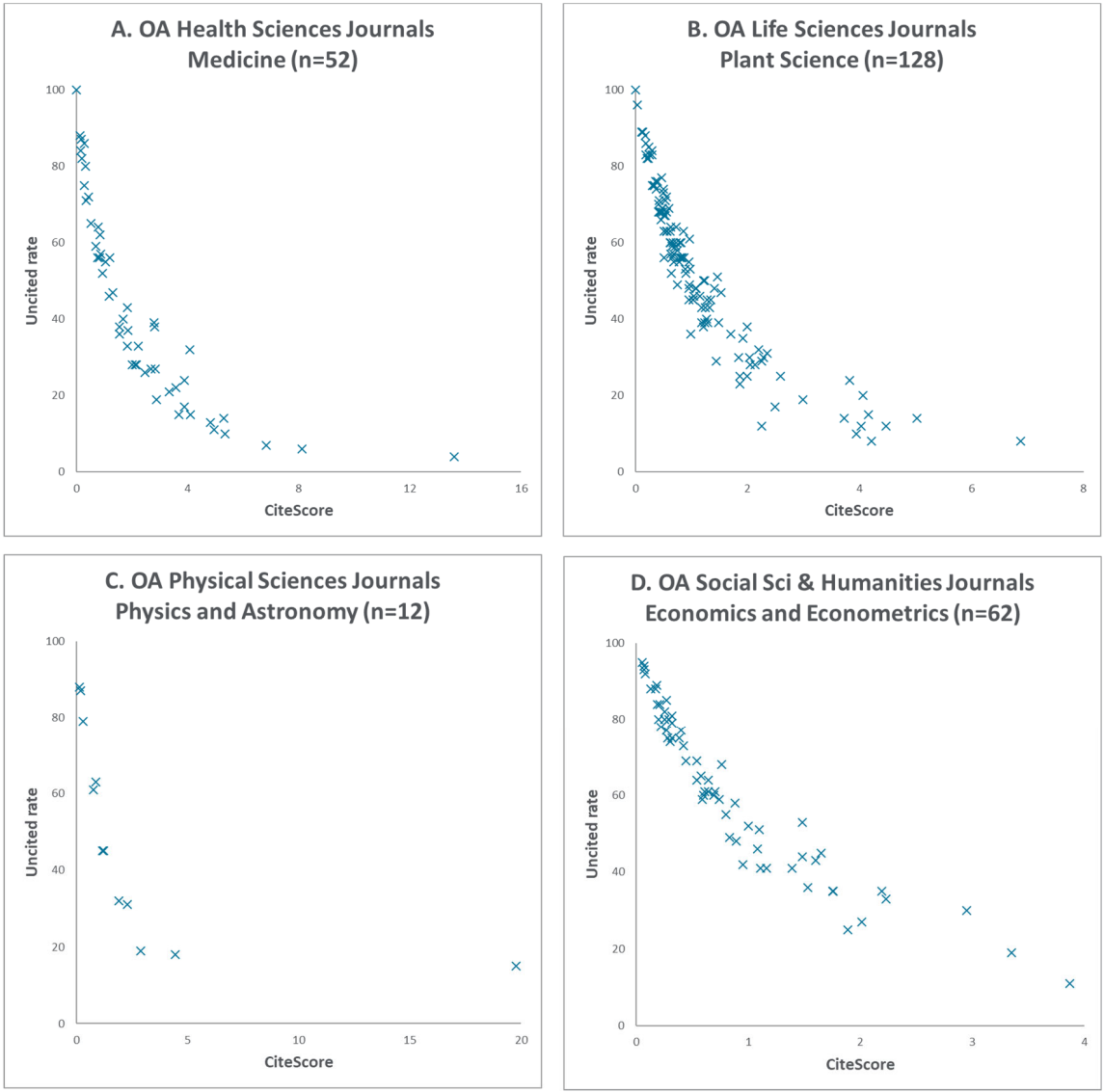

Figure B.2: Scatter plot between uncited rate and CiteScore for OA journals in four subject categories